\documentclass[aip,cha,reprint,nofootinbib]{revtex4-1}
\usepackage{amsfonts,amssymb,amsmath,graphicx,enumerate}
\usepackage{amsthm}
\usepackage{multirow}
\usepackage{times}
\usepackage{bm}
\usepackage{color,fancyref}

\begin{document}

\title{Disentangling regular and chaotic motion in  the standard map using complex network analysis of recurrences in phase space}

\author{Yong Zou}
 \affiliation{Department of Physics, East China Normal University, 200062 Shanghai, China}

\author{Reik V.~Donner}
\affiliation{Potsdam Institute for Climate Impact Research, P.\,O.~Box
60\,12\,03, 14412 Potsdam, Germany}

\author{Marco Thiel}
 \affiliation{Institute for Complex Systems and Mathematical Biology, University
 of Aberdeen, Aberdeen AB243UE, United Kingdom}

\author{J\"urgen Kurths}
 \affiliation{Potsdam Institute for Climate Impact Research, P.\,O.~Box
 60\,12\,03, 14412 Potsdam, Germany}
 \affiliation{Institute for Complex Systems and Mathematical Biology, University
 of Aberdeen, Aberdeen AB243UE, United Kingdom}
 \affiliation{Department of Physics, Humboldt University Berlin,
 Newtonstra{\ss}e 15, 12489 Berlin, Germany}
\affiliation{Department of Control Theory, Nizhny Novgorod State University,
Gagarin Avenue 23, 606950 Nizhny Novgorod, Russia}

\date{\today}

\begin{abstract}

Recurrence in the phase space of complex systems is a well-studied phenomenon,
which has provided deep insights into the nonlinear dynamics of such systems.
For dissipative systems, characteristics based on recurrence plots have recently
attracted much interest for discriminating qualitatively different types of
dynamics in terms of measures of complexity, dynamical invariants, or even
structural characteristics of the underlying attractor's geometry in phase
space. Here, we demonstrate  that the latter approach also provides a
corresponding distinction between different co-existing dynamical regimes of the
standard map, a paradigmatic example of a low-dimensional conservative system.
Specifically, we show that the recently developed approach of recurrence network
analysis provides potentially useful geometric characteristics distinguishing
between regular and chaotic orbits. We find that chaotic orbits in an
intermittent laminar phase (commonly referred to as sticky orbits) have a
distinct geometric structure possibly differing in a subtle way from those of
regular orbits, which is highlighted by different recurrence network properties
obtained from relatively short time series. Thus, this approach can help
discriminating regular orbits from laminar phases of chaotic ones, which
presents a persistent challenge to many existing chaos detection techniques.

\end{abstract}


\maketitle

\begin{quotation}

In recent years, complex network theory has provided many conceptual insights
based on recurrence characteristics of time series from various fields, which is
referred to as recurrence network analysis. While recent applications of this
novel concept have been restricted almost exclusively to dissipative dynamics
(i.e., the quantitative characterization of attractors), we demonstrate in this
work that some of the characteristic features of recurrence networks are useful
for  disentangling the complex dynamics of low-dimensional conservative systems
as well. In the standard map, a typical chaotic orbit  can be temporarily
trapped in the vicinity of  the regular domains in phase space, resulting in a
possibly rather long time necessary to homogeneously fill the chaotic domain --
a phenomenon known as stickiness. The presence of sticky orbits  (i.e.,
intermittent laminar phases of chaotic trajectories) presents an ongoing
challenge  to numerically characterizing the associated phase  portraits. In
this work, we demonstrate that in the standard map, the geometric organization
of  regular orbits as well as sticky  versus filling parts of chaotic orbits in
phase space can be successfully discriminated  based on relatively short time
series by using several  recurrence network measures, including network
transitivity, global clustering coefficient and average path length.  This
result provides the first documented finding pointing to the relevance of
recurrence network analysis for studying conservative dynamical systems.

\end{quotation}

\section{Introduction}
The importance of Poincar\'e's recurrences in dynamical systems has been widely
recognized ~\cite{Poincare1890}. In the last decades, considerable theoretical
progress has been made regarding the dynamical characteristics of various types
of complex systems. One particularly important achievement has been the
introduction of a rather simple visualization technique for recurrences in phase
space, the recurrence plot \cite{Eckmann1987,marwan2007}, which reduces the
fundamental complexity of studying recurrences to a binary matrix
representation. This conceptually simple mathematical form allows drawing upon
analogies to basic concepts of nonlinear time series analysis
\cite{Faure1998,thiel04}, information theory (statistics on binary sequences
providing measures of dynamical disorder and complexity)
\cite{zbilut92,Marwan2002herz} or, more recently, complex network theory
\cite{Marwan2009,Donner2010NJP,Donner2011IJBC}. These analogies have opened
important research avenues for using different types of statistics based on
recurrences, which are nowadays widely applied to time series from various
fields. While most recent studies have been restricted to dissipative dynamics
(i.e., the quantitative characterization of attractors), we demonstrate in this
work that some of the characteristic features of recurrence networks are useful
for  studying conservative systems as well.

The phase space of many non-integrable Hamiltonian systems is composed of
intermingled regions of regular and irregular orbits~\cite{Bountis2012}. The
regular  domain comprises the state vectors on both periodic and quasi-periodic
trajectories, while the irregular one contains  the states forming chaotic
orbits.  Here, a typical chaotic trajectory needs a  certain time to 
homogeneously fill its corresponding domain in phase space.  However, once a
chaotic orbit gets close to  a stable periodic island (i.e., the regular
domain), it can  be trapped in the vicinity of this domain and, hence, appear
almost regular in its motion for a  substantial amount of time. After this 
intermittent laminar phase, the orbit escapes again to the  rest of the chaotic
domain, eventually describing chaotic bursts across the termination of this
phase. Such  an intermittent, possibly long-term confinement of the trajectory
close to the regular domain is commonly referred to as stickiness
\cite{Karney_physicaD_1983,Meiss_rmp_1992}  and has been accepted as a
fundamental property of many Hamiltonian systems. Among other possible
scenarios~\cite{Afraimovich1998}, the existence of (regular)
islands-around-islands embedded in the chaotic domain is one of the mechanisms
that is able to generate stickiness \cite{Meiss_rmp_1992,Zaslavsky_phyrep_2002}.
{As a particularly relevant consequence, stickiness has been demonstrated to
result in anomalous transport phenomena in the phase space of Hamiltonian
systems.}

 The most traditional way of discriminating qualitatively different types of dynamics
is computing or numerically estimating the largest Lyapunov exponent
$\lambda$~\cite{Ott_1993}. Chaotic motion is characterized by positive $\lambda$
(in the case of  conservative systems, the sum of all exponents is zero). Regular
orbits, on the other hand, have zero Lyapunov exponents. In practice,
finite-time Lyapunov exponents are commonly used when resorting to numerical
calculations. In such cases, considerable attention  has to be paid to the
convergence rate  of the employed method, since the proper estimation of
$\lambda$ becomes extremely challenging if  the observed orbit encounters a sticky phase or exhibits another type of intermittent behavior~\cite{Afraimovich1998,Afraimovich2003}. Recent work~\cite{Kandrup_chaos_1999} has proposed circumventing the corresponding problems by considering sticky and filling chaotic phases separately. This distinction can be useful for
chaotic orbits that have a long sticking time \cite{Contopoulos_celesastron_1997}  and shall also be employed throughout the remainder of this manuscript.

 Beyond the concept of Lyapunov exponents, there is a vast body of methods for detecting chaos from time series~\cite{Bountis2012}. Most of these
approaches make use of the transverse stability of regular orbits in contrast to
the exponential divergence of initially close trajectories in the case of chaos,
thereby providing heuristic simplifications of the classical Lyapunov exponent
concept. As a potential alternative, in this work we  further explore the potentials of
recurrence plot-based methods (in particular, recurrence network  analysis) for
obtaining a discrimination between regular and chaotic trajectories from relatively short time series -- a problem where
other chaos indicators commonly experience difficulties.  Corresponding studies for
dissipative systems have already demonstrated the great potentials of such
approaches, but have not yet been systematically extended to  conservative
dynamics. However, the underlying concept of recurrences in phase space is
well-defined in both types of systems, so that it appears natural to apply
corresponding methods  to conservative systems as well. This work is intended to
fill this gap. 

As a paradigmatic example of an autonomous nearly-integrable system with two
degrees of freedom, we restrict our attention to the standard map
\begin{equation} \label{std_map_book}
\begin{aligned}
& y_{n+1} = y_{n} + \frac{\kappa}{2 \pi}\sin(2 \pi x_{n}),  \\
& x_{n+1} = x_{n} + y_{n+1},
\end{aligned} \;\;\text{mod}\;\; 1
\end{equation}
with $\kappa $ denoting the system's single control parameter, and
$\mathbf{v}_n=(x_n,y_n)$ being the state vector of the system at its $n$-th
iteration. This model is probably the best-studied chaotic Hamiltonian map and
can be interpreted as a Poincar\'e section of a periodically kicked rotor
\cite{Lichtenberg_Lieberman_regular,Meiss_rmp_1992}.

The remainder of this paper is organized as follows: Section~\ref{sec:methodS}
briefly reviews some of the mathematical concepts of recurrence analysis, comprising different approaches that address either
dynamical (R\'enyi entropy of second order $K_{2}$, mean recurrence time
$\left<RT\right>$) or geometric characteristics (recurrence network properties).
A more detailed description of all methods is provided as an Appendix. The
results obtained when applying the different methods to example trajectories
of the standard map are discussed in Section~\ref{sec:results}, providing some
particularly interesting findings based on the recurrence network approach. We
use this information for discussing the corresponding potentials to discriminate
between initial conditions yielding regular and chaotic orbits of the standard
map, focusing on the problem of  identifying and characterizing  chaotic orbits in their sticky phase.

\section{Methods} \label{sec:methodS}
Recurrence is a fundamental property of dynamical systems. In general, it can be
conveniently analyzed by means of \textit{recurrence plots}
(RPs)~\cite{marwan2007} originally introduced in the seminal work of Eckmann
\textit{et~al.}~\cite{Eckmann1987} This tool provides a two-dimensional
intuitive visualization of the underlying temporal recurrence patterns even for
high-dimensional systems. For this purpose, one defines the \textit{recurrence
matrix} (RM) $R_{i,j}$ as a binary representation of whether or not pairs of
observed state vectors on the same trajectory are mutually close in phase space.
Given two state vectors $\mathbf{v}_i$ and $\mathbf{v}_j$ (where $i$ and $j$
denote time indices), this proximity is most commonly characterized by comparing
the length of the difference vector between $\mathbf{v}_i$ and $\mathbf{v}_j$ to
a prescribed maximum distance $\varepsilon$, i.e.,
\begin{equation}
R_{i,j}(\varepsilon)=\Theta(\varepsilon-\|\mathbf{v}_i-\mathbf{v}_j \|),
\label{eq_defrp}
\end{equation}
where $\Theta(\cdot)$ is the Heaviside function and $\| \cdot\|$ a suitable
norm. In this work, we use the maximum norm for defining distances between state vectors in
phase space. The properties of RPs have been intensively studied for
different kinds of dynamics~\cite{marwan2007}, including periodic,
quasi-periodic~\cite{Zou_PRE2007,Zou_Chaos_2007}, chaotic, and stochastic
dynamics~\cite{marwan2007}.

The crucial parameter for the calculation of the RM is the recurrence threshold
$\varepsilon$. There are several rules-of-thumb to select a proper value of
$\varepsilon$. In many applications of RPs to time series from various fields,
it was found that the recurrence patterns do not change qualitatively for a
large range of $\varepsilon$, allowing for a reliable statistical analysis.
Furthermore, when comparing different time series (or trajectories), there are
two ways to set $\varepsilon$: (i) as a fixed value for all time series,
yielding possibly different values of the recurrence rate
\begin{equation}
RR=\frac{2}{N(N-1)}\sum_{i>j} R_{i,j}(\varepsilon),
\end{equation}
i.e., the density of non-zero entries in the RM, for different series; and (ii)
as a (variable) threshold value corresponding to a fixed (desired) $RR$. Note that the
method (ii) generally makes it easier to quantitatively compare the recurrence
structures of different time series, since some of the characteristic
statistical quantities based on the RM change systematically with an
increasing $RR$. In this work, we will revisit the difference between both
approaches for choosing $\varepsilon$ for a low-dimensional area-preserving map (Eq.
\eqref{std_map_book}).

There is a multiplicity of approaches allowing to characterize the dynamics of a
system under study based on its RM. First, it has been shown that, among other
features, the length of diagonal and vertical structures in RPs can be used for
defining a variety of complexity measures, which characterize properties such as
the degree of determinism or laminarity of the system \cite{zbilut92,trulla96}.
The resulting recurrence quantification analysis (RQA) has been widely applied
for studying phenomena from various scientific disciplines \cite{marwan2007}. As
demonstrated earlier~\cite{Zou_Chaos_2007}, RQA measures are able to
characterize the stickiness property of chaotic orbits in Hamiltonian systems like the standard map.
Therefore, in the present work, we do not include any further discussion of RQA
measures to avoid repetitions.

A second approach to use recurrence properties for a quantitative
characterization of the system's dynamics makes use of the fact that several
dynamical invariants, such as the R\'enyi entropy of second order $K_{2}$ and
the correlation dimension $D_{2}$, can be reliably estimated from the
RM~\cite{Faure1998,thiel03,thiel04}. Specifically, $K_2$ measures the average
rate at which information on previous states is lost during the system's
evolution. The inverse of the entropy value thus provides a rough estimate of
the time for which a reasonable prediction is possible. Accordingly, for a
sequence of independent and identically distributed random numbers (white
noise), $K_{2}$ tends to infinity, periodic dynamics is characterized by
$K_{2}=0$, and chaotic systems yield a positive yet finite $K_{2}$, as they
belong to a category between regular deterministic and stochastic systems in
terms of their predictability. A quasi-periodic system shows non-trivial
recurrences but low complexity \cite{Lichtenberg_Lieberman_regular}, which
yields $K_2 \approx 0$. Hence, $K_{2}$ is an appropriate measure to distinguish
qualitatively different behaviors of the system.  For the examples considered in
this paper, we use trajectories of $N=1,000$ to $5,000$ data points to
estimate $K_{2}$ for each orbit, applying the algorithm {proposed} by
Thiel \textit{et~al.}~\cite{thiel04}.

A third way to characterize the recurrence properties is statistically
evaluating the distribution of recurrence  times (RTs), which has been applied
to both chaotic and stochastic systems
\cite{Zaslavsky_phyrep_2002,altmannPRE2005}. Recurrence times refer to the time
intervals after which a trajectory enters the $\varepsilon$-neighborhood of a
previously visited point in phase space and are conveniently described by their
empirical probability distribution $p(\tau)$. Note that $p(\tau)$ contains
important information about the \textit{dynamics} of the system, which can be
used for detecting subtle dynamical transitions of the system under study as
some characteristic parameter is varied.  A periodic process shows a trivial RT
distribution that yields a delta-peaked $p(\tau)$ positioned at the system's
period.  A quasi-periodic process involving two incommensurate frequencies
(equivalent to a linear rotation on a unit circle with an irrational frequency)
displays three unique recurrence times, resulting in $p(\tau)$ being delta-peaked
at the three corresponding periods. Here, the largest characteristic RT is
simply the sum of the other two according to Slater's theorem
\cite{Slater_gaps_1967}. This theorem has been demonstrated to provide a useful
and fast tool for determining the presence of quasi-periodicity
\cite{Diego_pre_2006,Zou_PRE2007,Zou_Chaos_2007}. {Moreover,} RT statistics have great
potentials for estimating dynamical invariants (such as the information
dimension $D_1$ and the Kolmogorov-Sinai entropy \cite{BaptistaPRL2005}) and for
studying extreme events \cite{altmannPRE2005}. In turn, the recurrence \emph{of}
extreme events in dynamical systems has been recently discussed in terms of
generalized extreme value theory to define a dynamical stability indicator for
Hamiltonian maps \cite{Faranda2011}, an approach that is conceptually related to
the focus of this study, but shall not be further discussed here. An empirical
estimation of $p(\tau)$ from the RM is based on the same concept as other
recurrence  time statistics utilized elsewhere in the literature
\cite{Chirikov_physicaD_1984,Zaslavsky_phyrep_2002}, with a slightly different
averaging over all available state vectors $\mathbf{v}_i$ involved (see the Appendix
for details). The mean recurrence  time is calculated straightforwardly as
$\left< RT \right> = \int_{0}^{\infty} \tau p(\tau) d\tau$.

Finally, following a more recent approach, the RM (Eq.~\ref{eq_defrp}) can be
re-interpreted as the adjacency matrix of a complex network, the
$\varepsilon$-recurrence network (RN). 
In this context, each state vector $\mathbf{v}_i$ used in the computation of the RM is interpreted as a node of a complex network embedded in the phase space of the dynamical system under study.
The quantitative analysis of RNs
allows identifying transitions between different types of dynamics in a very
precise way \cite{Marwan2009,Donner2011IJBC,Zou2010,Donner2011EPJB}. In this
paper we consider three network measures that have already been shown to
distinguish between qualitatively different types of behavior in both discrete
and continuous-time dissipative systems
\cite{Marwan2009,Zou2010,Donges2011NPG,Donges2011PNAS}: \textit{global
clustering coefficient} $\mathcal{C}$, \textit{network transitivity}
$\mathcal{T}$, and \textit{average path length} $\mathcal{L}$ (see the Appendix
for details). Given the invariant density $\rho(x)$ of the system under study,
the value of $\mathcal{T}$ can be analytically computed~\cite{Donges2012PRE} and
interpreted as a generalized fractal dimension~\cite{Donner2011EPJB}.
Specifically, high values of $\mathcal{T}$ indicate the presence of a
lower-dimensional structure in phase space commonly corresponding to more
regular dynamics. In contrast, the average path length behaves differently for
different types of dissipative dynamical systems
\cite{Marwan2009,Donner2010NJP,Zou2010}: for maps, more regular dynamics is
characterized by lower values of $\mathcal{L}$, whereas the opposite applies to
continuous-time systems such as dissipative chaotic oscillators.

\begin{table}[htb]
\begin{tabular}{|l|l|}
\hline
Aspect  &  Measures     \\
\hline \hline
dynamic        & $K_2$,\ $\left< RT \right>$ \\
\hline
geometric     & $\mathcal{T}$,\ $\mathcal{C}$,\ $\mathcal{L}$ \\
\hline
\end{tabular}
\caption{All measures based on the RM used in this study. In addition, we
consider numerical estimates of the largest Lyapunov exponent $\lambda$ obtained
from considerably longer trajectory segments (see text) as benchmarks.}
\label{tab:measures}
\end{table}
Table~\ref{tab:measures} summarizes all measures that will be used in the
following. We note that the Lyapunov exponent $\lambda$, $K_2$ and the RT
distribution characterize time series from a dynamic perspective, whereas
recurrence network analysis discloses the properties of complex systems from a
geometric viewpoint. Certainly, the topological features of RNs are closely
related to dynamical characteristics of the underlying system
\cite{Donner2010NJP,Donner2011EPJB}.

\section{Results} \label{sec:results}

\subsection{Example trajectories: quasi-periodic, sticky and filling chaotic orbits} \label{sec3a}

We follow the approach of Kandrup \textit{et al.}~\cite{Kandrup_chaos_1999} and
categorize the trajectories of the standard map (Eq.~\ref{std_map_book}) into
quasi-periodic, sticky and filling (strongly) chaotic orbits. Note that
stickiness is a generic property of Hamiltonian chaos
\cite{Lichtenberg_Lieberman_regular} when $t \to \infty$. However, the concept
of a sticky orbit only refers to the particular time-scale during which it is
stuck (i.e., an intermittent laminar phase of the chaotic orbit's evolution). In this section, we use RN and RT characteristics to investigate whether or not it is possible to distinguish between
quasi-periodic orbits and the sticky (laminar) phases of chaotic orbits from relatively short trajectory segments. Regarding other
measures of RQA, we refer to our previous work~\cite{Zou_Chaos_2007}.

\begin{figure}
	\centering
	\includegraphics[width=\columnwidth]{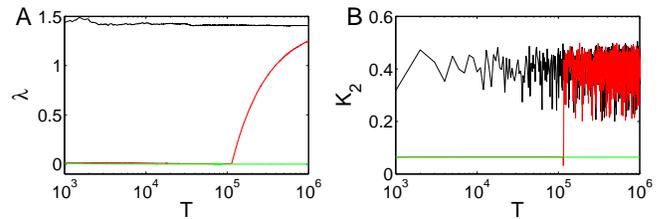}
\caption{\small{(color online) (A) Largest Lyapunov exponent $\lambda$ for three selected orbits ($\kappa = 5$ in Eq. \ref{std_map_book}) in dependence on the considered number of iterations. The different lines represent quasi-periodic (green), sticky (red), and filling chaotic (black) orbits. (B) R\'enyi entropy $K_2$ computed for non-overlapping windows sliding along the trajectories (window length $N=1,000$). } \label{lek2_stick}}
\end{figure}

To illustrate our approach, we first consider three typical segments of orbits of the standard map~\cite{Contopoulos_celesastron_1997}, one of which shows
the stickiness property clearly. In Fig.~\ref{lek2_stick}A, numerical estimates
of $\lambda$ are shown for these three example segments. Here, the initially sticky chaotic trajectory segment
escapes from the vicinity of the regular domain after approximately $1.65 \times 10^5$
iterations. We call this the escape time $T_{esc}$, which is about two orders of
magnitude larger than the common ``observation'' period in many time series
analysis applications ($10^3$ to $10^4$ time steps). For better comparison, we
also choose a quasi-periodic orbit together with a filling part of a chaotic one. We note
that with the number of iterations being much smaller than $T_{esc}$, it is not
possible to distinguish the quasi-periodic orbit from the sticky part of the chaotic trajectory based on
numerical estimates of $\lambda$. Furthermore, the entropy $K_2$ also reveals rather
little difference between the sticky segment and the quasi-periodic orbit if the number of
iterations does not reach $T_{esc}$ (cf.\, the red and green lines in
Fig.~\ref{lek2_stick}B). These results demonstrate that it is necessary to look
at other properties that allow for a corresponding discrimination
already from time series that are considerably shorter than $T_{esc}$.

\begin{figure}
	\centering
	\includegraphics[width=\columnwidth]{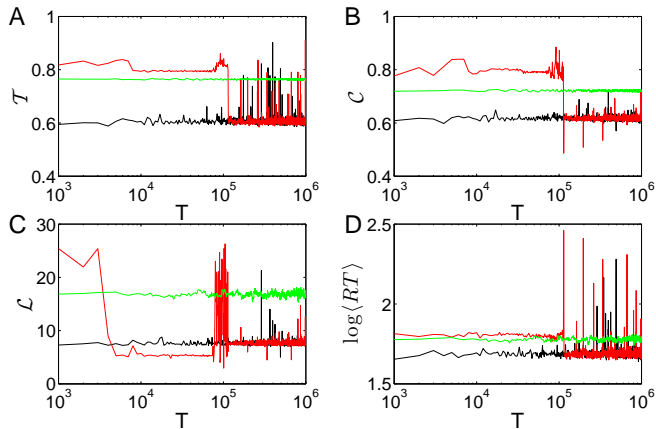}
\caption{\small{(color online) Recurrence characteristics for the quasi-periodic
(green), sticky (red), and filling chaotic (black) orbits from
Fig.~\ref{lek2_stick} computed for sliding windows ($N=1,000$) over the
different time series: (A) RN transitivity $\mathcal{T}$; (B) global clustering
coefficient $\mathcal{C}$; (C) average path length $\mathcal{L}$; (D) logarithm
of the mean recurrence time $\left<RT\right>$. }\label{netM_stick}}
\end{figure}

Figure~\ref{netM_stick} shows the behavior of the three RN-based measures
$\mathcal{T}, \mathcal{C}$, and $\mathcal{L}$, together with the mean recurrence
times $\left<RT\right>$ for the three considered trajectory segments,
respectively. In all computations, we have used $RR=0.02$, maximum norm and
sliding windows with $N=1,000$ subsequent state vectors taken from the respective
orbits. We note that a careful statistical analysis of the
probability distribution functions of all four considered measures for different
types of dynamics would be required for evaluating whether or not statistically
significant deviations exist among different types of trajectories. However,
such a detailed analysis is beyond the scope of the present work. To this end,
we emphasize that by visual inspection, one can clearly recognize that all four
measures reveal distinctive differences between the considered quasi-periodic
and filling chaotic  orbit segments. In contrast, the sticky example orbit
exhibits some distinct and quite complex behavior during different time
intervals. In the latter case, the values of all considered measures quickly converge towards those of the filling chaotic orbit after the termination of the sticky phase, i.e., after the trajectory has escaped from the vicinity of the regular
domain in phase space. However, a more detailed inspection reveals that based upon the RN characteristics we can distinguish three different stages: $T<T_{dec}$, $T_{dec}<T<T_{esc}$ and
$T>T_{esc}$. We will discuss the meaning of $T_{dec}$ ($\approx 4\times
10^3$) below in some detail. The first two stages both correspond to the sticky phase of the orbit and can only be distinguished from each other by the observed values of
$\mathcal{L}$, whereas the third one corresponds to the post-escape phase. The latter part of the trajectory comprises a phase of bursting behavior characterized by huge variations of all considered characteristics, which is followed by a more homogeneously filling chaotic phase interrupted by just a few short bursting intervals possibly indicating further short periods of intermittent dynamics.

In the post-escape phase ($T>T_{esc}$), the chaotic orbit has left the vicinity of the regular domain (i.e., its intermittent laminar phase) and fills the chaotic domain of the phase space in essentially the same way as the initially filling chaotic trajectory segment. Strong fluctuations of the considered recurrence characteristics in the first part of this phase indicate the presence of chaotic bursts where the geometric structure of the orbit observed for short periods of time does not yet correspond to that of the homogeneously filling chaotic dynamics, but exhibits a strong variability. In turn, in the homogeneously filling chaotic regime, previous results for dissipative maps suggest that the three RN
characteristics $\mathcal{T}$, $\mathcal{C}$ and $\mathcal{L}$ should consistently show much larger values for the regular (quasi-periodic) than for the chaotic orbits (with the exception of intermittent bursts possibly representing short periods of stickiness), which is confirmed by our numerical results. 

We emphasize that intermittent bursts with strongly fluctuating RN
and RT characteristics as observed in Fig.~\ref{netM_stick} are characteristic
for chaotic trajectory segments and cannot be observed in the case of
quasi-periodic dynamics. Hence, the presence of abrupt changes in the recurrence
characteristics could serve as a criterion for identifying chaos.
Notably, the latter viewpoint is not helpful when experiencing a
sticky phase -- or another type of intermittent behavior -- during the entire
period of available observations. However, the probability to
encounter such a situation when {choosing} the initial conditions at random
decreases systematically as the length of the studied trajectory segment
increases.

Considering the laminar phase ($T<T_{esc}$), we find that all four
characteristics show different values than for the regular (quasi-periodic) one.
Notably, $\mathcal{T}$, $\mathcal{C}$ and $\left<RT\right>$ are substantially
larger during the sticky phase of the chaotic trajectory in
comparison with the regular orbit, while the corresponding
difference is most clearly visible in the two RN characteristics
(Fig.~\ref{netM_stick}D). At first, this result is surprising, since
$\mathcal{T}$ and $\mathcal{C}$ have an interpretation as the global and
spatially averaged local (geometric) dimension of the trajectory,
respectively \cite{Donner2011EPJB}. Hence, one could expect that
the values of both measures are smaller for a chaotic orbit (even in
a sticky phase) than for a quasi-periodic one. In turn, based upon numerical
studies, Tsiganis \textit{et~al.} \cite{Tsiganis2000} argued that ``the
dimension of the phase space subset on which a sticky segment is embedded does
not differ from the dimension of the set on which a regular orbit lies.''
According to these results, at least the correlation dimension studied in the
former work allows discriminating between sticky and filling chaotic phases, but
not between sticky chaotic segments and a quasi-periodic orbit.

\begin{figure}
	\centering
	\includegraphics[width=\columnwidth]{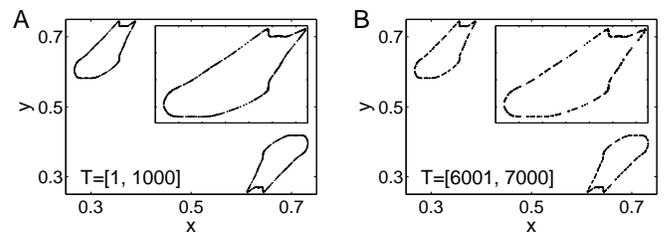}
\caption{\small{Phase portrait of the chaotic orbit in its sticky phase} during two different periods of
iterations: (A) $T = [1, 1000]$, (B) $T = [6001, 7000]$. Inserts show a zoom of
one part of the orbit, displaying the different arrangements of the associated
sets of state vectors in phase space (different distribution of gaps between
neighboring state vectors). \label{stick_phase}}
\end{figure}

In order to understand why we can actually observe a marked
difference between sticky segments and neighboring quasi-periodic orbits,
Fig.~\ref{stick_phase} highlights the geometry of the set of state vectors
forming the trajectory segment during the sticky phase. We observe that as the
orbits within the nearby quasi-periodic domain, in its sticky
phase the chaotic orbit consists of two spatially separated
parts, each of which displays two sharp kinks. For the
dissipative chaotic H\'enon map, it was found that such structures
can spuriously induce a lower dimensionality of the system (for
finite values of $\varepsilon$) due to the existence of additional
closed triangles in the RN in the parts of the phase space where
these kinks are located and, hence, results in a positive bias of both
$\mathcal{T}$ and $\mathcal{C}$~\cite{Donner2010NJP,Donner2011EPJB}.

Regarding the average path length $\mathcal{L}$, Fig.~\ref{netM_stick} shows
that the RN is initially ($T<T_{dec}$) characterized by higher values for the
chaotic orbit in its laminar phase in comparison with the quasi-periodic one, whereas $\mathcal{L}$ decays fastly around $T=T_{dec}$ and reaches values even below those found for the initially filling chaotic
orbit for $T>T_{dec}$. The reason for this marked drop can {also} be
identified from Fig.~\ref{stick_phase}. Initially, the orbit appears more or less like a
(discretely sampled) continuous curve (i.e., spatially neighboring state vectors
are always separated by distances less than $\varepsilon$). In turn, at
$T>T_{dec}$ distinct subsets of state vectors spanning the orbit segment get separated by increasingly large
phase space regions (larger than $\varepsilon$) without state
vectors. Hence, we refer to $T_{dec}$ as the RN decomposition time, which depends on the chosen value of $\varepsilon$. Consequently, the initial
RN consists of mainly two mutually disconnected network components
corresponding to two spatially separated parts of the regular domain in phase space
(Fig.~\ref{stick_phase}A), but then decomposes further into a larger number of
mutually disconnected network components (which are mutually separated by more than a distance
$\varepsilon$ in phase space) at $T>T_{dec}$ ({cf.\, the inset of}
Fig.~\ref{stick_phase}B). Since the average path length is commonly computed
over all pairs of mutually reachable nodes, it necessarily decreases if there
is a transition towards more components of smaller size.

\begin{figure}
	\centering
	\includegraphics[width=\columnwidth]{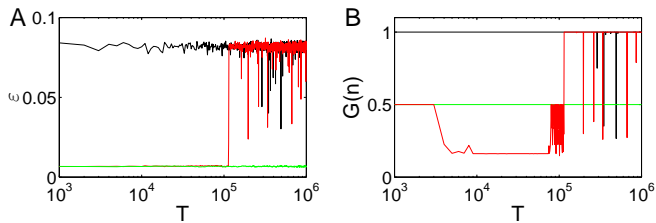}
\caption{ \small{(A) Recurrence threshold $\varepsilon$ and (B) size of the RN's
largest component $G(n)$ for sliding windows of $N=1,000$ points along the three
previously considered trajectory segments. Note that due to the presence of two
spatially separated parts of the orbits, the quasi-periodic orbit (green) and the sticky phase of the chaotic
orbit} (red) have $G(n)\leq 0.5$, whereas the filling chaotic one (black)
commonly exhibits a single component (exceptions likely correspond to short
periods of stickiness or effects due to the finite sample size), i.e.,
$G(n)=1$.\label{components}}
\end{figure}

The latter explanation is confirmed by a sharp decrease in the size of the RN's
giant component shown in Fig.~\ref{components}B. We emphasize that the number of
RN components $N_c$ would provide another suitable measure for
tracing this transition. In turn, in the given situation, the RN percolation
threshold $\varepsilon_c$ is less suited for this purpose, since the regular
domain close to which the sticky orbit segment is located already
consists of two spatially separated parts (as seen from the size of
the largest RN component for the nearby quasi-periodic orbit exhibiting
$G(n)\leq 0.5$). In contrast, the RN properties $\mathcal{L}$ and $G(n)$ unveil
the mechanism how the laminar phase of the chaotic orbit is
terminated: the state vectors along the chaotic trajectory get more
and more concentrated in phase space, leading to the successive
decomposition of the corresponding RN structure into smaller,
mutually disconnected subnetworks when  viewed at finite spatial
resolution $\varepsilon$ for finite time windows far before the actual escape at
$T=T_{esc}$. In this spirit, the decomposition of the RN provides an early
indication of the {loss of the} stickiness property of {the observed chaotic}
orbit. While the observations described above correspond to purely
empirical findings, an in-depth analysis of the intermittent behavior of chaotic
orbits of the standard map by complementary techniques like first-return maps
would allow relating the reorganization of chaotic orbit segments across the
escape from the sticky phase to the associated transport properties. Obviously,
along with the loss of stickiness the vicinity of the invariant tori are visited
more and more heterogeneously. The detailed examination of the dynamical roots
of this behavior (in the context of existing results on intermittency in
dynamical systems) will provide a potentially interesting topic for further
studies. The same applies to the processes accompanying the trapping of a
chaotic trajectory near periodic islands.

Within the framework of the present work, it would be a practically relevant question whether the observations described
above are also characteristic for sticky orbits in other Hamiltonian systems.
However, a detailed study of this aspect is beyond the scope of the present
work. To this end, we emphasize that systematically studying $T_{dec}$ as a
function of $\varepsilon$ could provide a tool for further quantitatively
characterizing this transition. However, it needs to be kept in mind that the RN
computed for running windows in time (i.e., consisting of a fixed number of
nodes) needs to have a minimum number of edges (minimum feasible $RR$
\cite{Donges2012PRE}) in order to allow for a proper evaluation, which sets a practical
lower bound to $\varepsilon$.

Regarding the numerical results described above, it only
remains to be explained why the average path length $\mathcal{L}$ initially
shows larger values during the sticky phase of the chaotic
trajectory segment than for the quasi-periodic orbit. Given that the recurrence
threshold $\varepsilon$ is almost the same for both orbits during the full
period of stickiness (cf.~Fig.~\ref{components}A), we relate this to the fact
that (unlike for \emph{dissipative} maps previously studied elsewhere
\cite{Donner2010NJP,Donner2011EPJB}) both the regular (quasi-periodic)
orbit and the sticky phase of the chaotic one appear (up to the
{spatial} resolution $\varepsilon$) to correspond to subsets of
state vectors that exhibit no gaps larger than $\varepsilon$ in phase space
(i.e., there exist relatively long paths in the RN connecting state
vectors with a mutual distance of less than $\varepsilon$). In such a case, the
behavior of $\mathcal{L}$ { is mainly determined by the lengths of these
paths, which can be analytically computed if the probability
distribution function of the state vectors is known~\cite{Donges2012PRE}. For
the sticky phase of the chaotic orbit, the
corresponding state vectors form an outer envelope of} the
quasi-periodic domain, so that the resulting larger spatial
dimensions imply somewhat larger $\mathcal{L}$ in the RN.

Summarizing the results obtained so far, all four characteristics
can potentially distinguish between quasi-periodic and sticky chaotic motion.
In particular, for $\mathcal{T}$ and $\mathcal{C}$ (but to a smaller
extend also $\left<RT\right>$), the observed values are consistent for the full
period of stickiness and thus allow identifying sticky orbit segments even from
relatively short time series (i.e., when the number of iterations is much lower
than $T_{esc}$, e.g., $10^3-10^4$ iterations), because the geometric shapes of
both types of orbits differ markedly. However, the aforementioned measures
over-compensate the relative tendency one would actually expect for chaotic
trajectories when compared to quasi-periodic ones. The latter observation is due
to the stronger ``kinkiness'' of the sticky orbits. We will re-examine this feature below
in order to address the question whether it can be exploited for automatically
discriminating between quasi-periodic and sticky chaotic trajectory segments.

\subsection{Full phase space} \label{sec3b}

Based on the results for the three example trajectories described in Section~\ref{sec3a}, we now
turn to a characterization of the full phase space regarding different types of
dynamics. For this purpose, we start with 500 initial conditions distributed
randomly within the domain of definition of the standard map, $(x,y) \in [0, 1]
\times [0, 1]$. Here, we use the canonical parameter value of $\kappa = 1.4$ in
Eq.~\ref{std_map_book}. All trajectories are computed for $5000$ time steps.
Note that for conservative dynamics we do not have transients in the
sense of dissipative dynamics where an attractor needs to be approached first.
This is an advantage in comparison to numerical studies of dissipative systems.
Hence, the state vectors at \emph{all} iterations can be further taken into
account.

\begin{figure}
	\centering
	\includegraphics[width=\columnwidth]{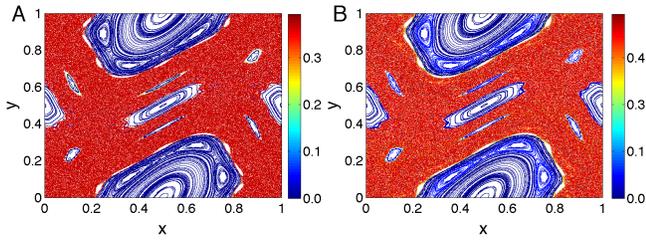}
\caption{\small {(color online) Phase space of the standard map
(Eq.~\eqref{std_map_book}) characterized by (A) the largest Lyapunov exponent
$\lambda$ and (B) $K_2$ estimated from the RM. } \label{fig:sm_le}}
\end{figure}

For the ease of comparison, Fig.~\ref{fig:sm_le}A displays the phase space of
the standard map, where a sample of state vectors of the system is colored according to the largest Lyapunov
exponent $\lambda$ of the orbit traversing each corresponding point in phase space. As expected, we observe a
significant difference between the chaotic domain with positive $\lambda >0$
and the regular domain with $\lambda\approx 0$. Note that for numerically
approximating $\lambda$ (but not the characteristics estimated from the RM), we
have used much longer trajectories with $N=2\times 10^5$ points, since the $N=5,000$
iterations considered in the following for computing the recurrence-based
characteristics {are often not} sufficient for obtaining stable estimates of
$\lambda$.

{The resulting pattern of} $K_2$ estimated from the RM is shown in Fig.~\ref{fig:sm_le}B.
Regarding our ``experimental'' setting, we emphasize that 5000 iterations often
do not guarantee that \emph{all} sticky orbits can leave the vicinity of the
quasi-periodic areas in phase space and, hence, that a single chaotic trajectory
segment can fully cover the chaotic domain. Therefore the considered trajectory
length of $N = 5,000$ state vectors seems to be insufficient to reliably estimate $K_2$, since
the scaling region used for estimation~\cite{thiel2004a} can eventually get
blurred by short diagonal lines. {This relatively short time series length explains that the numerical results of Fig.~\ref{fig:sm_le}B do} not
show convincingly the theoretical relationship that $K_2$ is a lower bound of
the sum of positive Lyapunov exponents of the system~\cite{Ott_1993}. However,
our ensemble of 500 random initial conditions is found to be sufficient to
{provide a rough map of} the phase space (distinguishing qualitatively different types of dynamics and highlighting domains with temporary stickiness of chaotic orbits) as shown below. Taken together, quasi-periodic orbits and
chaotic trajectories in their sticky phases cannot be {reliably} distinguished by {$K_2$ in the considered analysis setting}.

\begin{figure}
	\centering
	\includegraphics[width=\columnwidth]{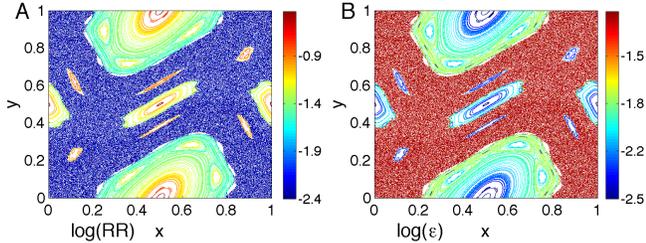}
\caption{\small {(color online) Same {phase space as in} Fig. \ref{fig:sm_le}, but {highlighting the two different} ways to choose the recurrence threshold $\varepsilon$. (A) $RR$ for a
fixed threshold $\varepsilon=0.03$, (B) $\varepsilon$ for a fixed $RR=0.02$. }
\label{fig:rec_rr}}
\end{figure}

As discussed in detail in Section~\ref{sec:methodS}, there are two main strategies
for selecting the threshold $\varepsilon$. While the estimation of $K_2$
(Fig.~\ref{fig:sm_le}B) from the RM does not distinguish between both approaches
(since it uses a scaling relationship emerging as $\varepsilon$ is varied), the
results for the other four characteristics depend on whether $\varepsilon$ or
$RR$ is fixed. Figure~\ref{fig:rec_rr} illustrates the mutual dependence between
$\varepsilon$ and $RR$, thereby extending upon earlier results for a single orbit
previously reported for dissipative chaotic systems~\cite{Donner2010PRE}. Note that for Hamiltonian {systems},
domains covered by periodic, quasi-periodic and chaotic orbits {can} have
intrinsically different sizes, hence, distances along such orbits are commonly
not comparable. Hence, using the same $\varepsilon$ can lead to very different $RR$
(Fig.~\ref{fig:rec_rr}B).

\begin{figure}
	\centering
	\includegraphics[width=\columnwidth]{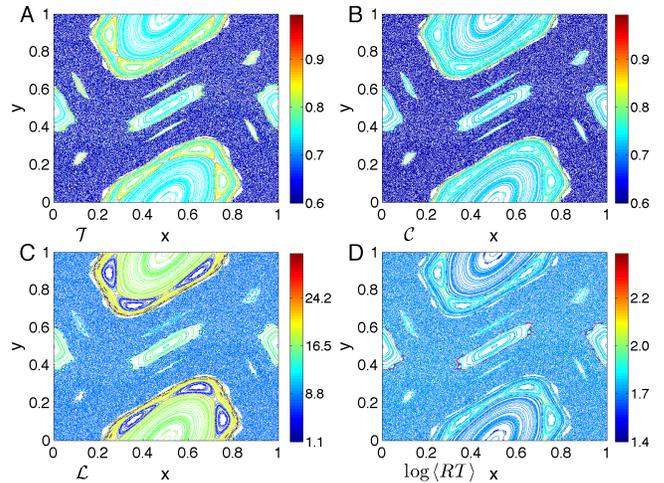}
\caption{\small {(color online) Same {phase space as in} Fig. \ref{fig:sm_le}, but characterized
by recurrence statistics for the standard map using fixed $RR=0.02$. (A)
$\mathcal{T}$, (B) $\mathcal{C}$, (C) $\mathcal{L}$, and (D) $\left<RT\right>$.}
\label{fig:sm_rec_rr}}
\end{figure}

\begin{figure}
	\centering
	\includegraphics[width=\columnwidth]{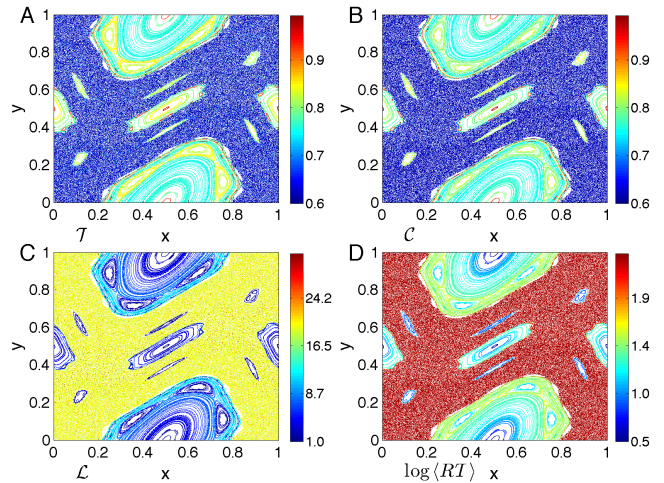}
\caption{\small {(color online) Same {phase space as in}  Fig.~\ref{fig:sm_rec_rr} for a fixed
$\varepsilon=0.03$. }
\label{fig:sm_rec_eps}}
\end{figure}

When aiming for a quantitative comparability of RN characteristics (which can
depend on $RR$), we suggest to adaptively choose $\varepsilon$ such that the
$RR$ has the same fixed value (Fig.~\ref{fig:rec_rr}B). In turn, since regular
and (even sticky) chaotic orbits can have considerably different spatial
dimensions, it is of potential interest to also consider a setting with $\varepsilon$
being globally fixed (see Fig.~\ref{fig:rec_rr}A for the resulting behavior of
$RR$). The corresponding results for the three RN measures as well as
$\left<RT\right>$ for both settings are shown in Figs.~\ref{fig:sm_rec_rr} and
\ref{fig:sm_rec_eps}, respectively.

The overall structure of the phase space with its intermingled regular and chaotic orbits is captured by both dynamic ($K_2$ and $RT$) and geometric measures
(e.g., $\mathcal{T}$ and other RN characteristics), as shown in Figs.
\ref{fig:sm_rec_rr} and \ref{fig:sm_rec_eps}. Specifically, quasi-periodic
trajectories are characterized by {larger} values of network transitivity
$\mathcal{T}$, while filling chaotic ones have smaller $\mathcal{T}$. The same
applies to $\mathcal{C}$. Regarding both properties, the obtained picture is
consistent for the two settings with fixed $RR$ and fixed $\varepsilon$,
respectively. For $\left<RT\right>$, the pattern is only conclusive for fixed
$\varepsilon$, where the mean recurrence times are considerably larger in the
filling chaotic case than for regular orbits (as it is the case for the RN
average path length $\mathcal{L}$), which is expected since the chaotic domain is larger than the regular one. Specifically, while a chaotic orbit can fill the complete domain (as $t\to\infty$), regular ones are distinct and mutually nested, which results in different values of $\left<RT\right>$ and $\mathcal{L}$ for the latter which (for fixed $\varepsilon$) depend clearly on the size of the orbit. {In turn, when fixing $RR$ the effect of different spatial distances on the estimated recurrence times and RN average path lengths is essentially removed (Fig.~\ref{fig:sm_rec_rr}C,D).}

\begin{figure}
	\centering
	\includegraphics[width=\columnwidth]{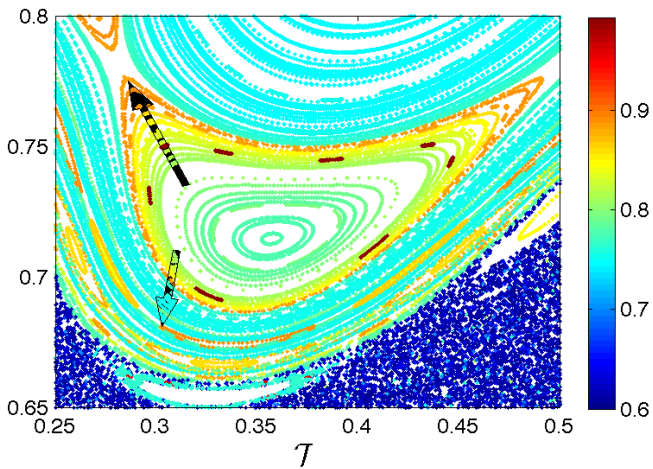}
\caption{\small {(color online) Zoom into Fig.~\ref{fig:sm_rec_rr}A displaying
the values of $\mathcal{T}$ at the interface between regular and chaotic
domains. Black arrows indicate chaotic areas with increased transitivity values.
}
\label{fig:zoom}}
\end{figure}

Turning back to the original question of {whether or not it is possible} to distinguish chaotic trajectories in their sticky phase from quasi-periodic orbits, our results indicate that in agreement with the
findings for the three example trajectories in Section \ref{sec:results}.A, the
RN transitivity $\mathcal{T}$ obtained with a fixed $RR$ is a promising candidate measure. Notably, {the} sticky chaotic orbit segments are organized
in phase space like envelopes of the islands-around-islands (i.e., a period-$n$ orbit surrounded by
islands of high periods). As Fig.~\ref{fig:zoom}
indicates, these envelopes are in fact characterized by elevated values of
$\mathcal{T}$ clearly above those found for the orbits within the regular
domains ($\mathcal{T}=0.75$ {-- as expected for one-dimensional curves~\cite{Donner2011EPJB} ---} in the middle of these domains and
$\mathcal{T}\approx 0.8$ inside the quasi-periodic island chains), while many
initial conditions close to the quasi-periodic orbits (but not belonging to
them) lead to finite-time trajectory segments with $\mathcal{T}\sim 0.85\dots 0.9$ and even above due to the kinky geometry shown in Fig.~\ref{stick_phase}{, which is consistent with previous findings for dissipative maps~\cite{Donner2010NJP,Donner2011EPJB}}.

Following the ideas presented in our previous work on dissipative
systems~\cite{Zou2010}, one could in principle assess the quality of a
classification obtained by the approach presented above in a more rigorous
statistical framework with the help of the largest Lyapunov exponent $\lambda$.
However, such an assessment of the discriminatory power of different measures
becomes both numerically and conceptually more challenging than for the problem of distinguishing
periodic and chaotic orbits {in the \emph{parameter} space of dissipative} nonlinear oscillators as
previously {studied}~\cite{Zou2010} for different reasons: (i) there are periodic,
quasi-periodic and chaotic trajectories (i.e., three qualitatively different types of orbits)
intermingled in phase space; (ii) there is no unique residence time scale for
chaotic orbits to stay in an intermittent laminar phase corresponding to stickiness; and (iii) technically,
the assessment is more appropriate for investigating the parameter space of
complex systems~\cite{Zou2010} than for exploring the phase space with
coexisting domains of qualitatively different dynamics as done as in Figs.
\ref{fig:sm_rec_rr} and \ref{fig:sm_rec_eps}. The statistical analysis would
hence yield results that are specific to the particular choices of the length of
time series and initial conditions.

\section{Conclusions}

We have presented a numerical case study demonstrating that
recurrence-based methods in general, and recurrence networks in particular, can
be useful for disentangling different dynamical regimes not only in dissipative,
but also in conservative dynamical systems.
Specifically, we have shown that the geometric properties of regular
(i.e., periodic and quasi-periodic) orbits of the standard map as captured by
recurrence networks clearly differ from such of (filling) segments of chaotic
orbits that do not exhibit intermittent periods of stickiness close to the
domain of regular solutions. While in the latter case, many existing approaches
can fail to unambiguously discriminate between quasi-periodic and sticky chaotic
dynamics, recurrence network characteristics provide indicators for the presence
of chaos even in such rather complex situations. The transition phase between
laminar (sticky) and homogeneously filling chaotic dynamics has not been
explicitly studied, but presents an interesting subject to be further
investigated. In general, one has to note that sticky and filling chaotic
dynamics can correspond to different phases of the same chaotic orbit (the
durations of which can show a considerably wide distribution). Thus, making a
distinction between both appearances is only reasonable for specified periods of
observation of the system under study.

It is important to emphasize that the proposed application of
recurrence network analysis to conservative systems has been successful already
for relatively short trajectory segments {(e.g., $N=1,000$ or $5,000$ points)}.
Although this study has considered only the paradigmatic and best-studied
example of a discrete-time two-dimensional Hamiltonian system exhibiting
co-existence between regular orbits and chaos (the Chirikov standard map), it is
expected that similar results can be obtained for other {Hamiltonian} maps with
a chaotic regime, such as a four-dimensional version of the standard map
consisting of two coupled classical standard maps~\cite{Freistetter2000},
Zaslavsky's stochastic web map~\cite{Zaslavsky1998}, Meiss' quadratic
map~\cite{Meiss1986}, or even continuous-time Hamiltonian system like the
H\'enon-Heiles system~\cite{Henon_Heiles_astro_1964,Zou_PRE2007}. A continuation
of this study taking such different systems into account will be subject of our
future work.

The aim of the present study was to initially explore the general potentials of
recurrence network analysis for applications to conservative
systems. We did neither undertake an exhaustive quantitative comparison between
different recurrence-based techniques, nor between recurrence-based approaches
and classical chaos indicators based on the idea of exponential transverse
motion in phase space {\cite{GottwaldChaos2014,Bountis2012} or established tests
for the presence of chaos like the 0-1 test~\cite{Gottwald2004}.} However, a
corresponding in-depth investigation should be a subject of future research as
well, including a systematic study of the effects of the available time series
length on the obtained estimates.

In order to further support the applicability of the proposed approach, a
detailed performance assessment of different methods would be desirable. For
such an assessment, two criteria appear of special relevance:

On the one hand, the classification accuracy needs to be addressed. Notably, a
corresponding investigation would require considerable numerical efforts, since
the exact location of the chaotic domains cannot be computed analytically for
the standard map, but needs to be evaluated by some benchmark technique. For
this purpose, the standard reference would again be the largest Lyapunov
exponent, the appropriate estimation of which, however, is considerably
challenged by the presence of stickiness at all possible time scales (i.e., the
closer an initial condition is to the boundary of the regular
domain, the longer the sticking time can be expected to be). In fact, this
problem affects essentially all existing classification criteria for orbits in
Hamiltonian systems exhibiting stickiness phenomena.

On the other hand, convergence time should be taken into account as a second
criterion, which is of particular relevance in situations where only relatively
short time series are available for characterizing the nature of different
orbits. Regarding the latter aspect, we emphasize that recurrence methods
present a powerful methodological alternative to indicators based on transverse
expansion. From the computational perspective, recurrence plot-based methods are
conceptually simple and have reasonable computational demands, making them
excellent candidates for applications to short experimental time series~\cite{marwan2007}.

\begin{acknowledgments}
YZ acknowledges financial support by the National Natural Science Foundation of
China (Grant Nos.~11305062, 11135001, 81471651), the Specialized Research Fund
(SRF) for the Doctoral Program (20130076120003), and the SRF for ROCS, SEM. RVD
has been supported by the Federal Ministry for Education and Research (BMBF) via
the young investigators group CoSy-CC$^2$ (project no. 01LN1306A).
\end{acknowledgments}

\appendix
\section{Methodological details} 

In this Appendix, we provide algorithmic details on the set of complementary
recurrence analysis approaches used in this paper.

\subsection{Dynamical invariant $K_2$}
We recall the techniques presented in~\citet{thiel03,thiel04} to estimate $K_2$
from the RM (Eq.~\ref{eq_defrp}) and present some remarks on the corresponding
computations. $K_{2}$ can be estimated from the cumulative distribution of
diagonal lines $P^c_{\varepsilon}(l)$ in the RP~\cite{thiel03}.
Specifically, the probability of finding a diagonal line of at least length $l$
in the RP of a chaotic map is given by
\begin{equation}\label{cumulative}
P^c_{\varepsilon}(l)\sim \varepsilon^{D_{2}}\exp({-K_{2}(\varepsilon) l)}.
\end{equation}

Therefore, if we represent $P^c_{\varepsilon}(l)$ in a logarithmic scale versus
$l$, we should obtain a straight line with slope $-K_{2}(\varepsilon)$ for
large $l$. This slope is independent of $\varepsilon$ for a deterministic
chaotic system, while linearly dependent on $\varepsilon$ for random processes.
Thus, $K_{2}$ can be estimated from RPs provided the length of time series
covers the underlying system in the (sampled) phase space sufficiently well. This
method has been successfully applied to characterize fluid dynamics in different
regimes~\cite{thiel04}, to study the stability of extrasolar planetary
systems~\cite{Planetary}, and to divide the parameter space of a mechanical
oscillator system into different regimes \cite{Zou2005}.

One important advantage of the RP-based estimator of $K_{2}$
(Eq.~\ref{cumulative}) is that it is independent of the choice of the parameters
for a possibly necessary embedding{, which can be} important when studying real-world observational time series.

\subsection{Recurrence time statistics from RM}

The detailed steps to estimate the RT distribution $p(\tau)$ from the RM are as
follows:

RTs can be identified as the lengths of non-interrupted vertical (or horizontal,
since the RM is symmetric) ``white lines'' that do not contain any recurrence
(i.e., no pair of mutually close state vectors). More precisely, such a white
line of length $\tau$ starts at the position $(i,j)$ in the RP if~\cite{thiel03}
\begin{equation} \label{wvl_eq}
R_{i, j+m} = \begin{cases}
 1 & \text {if} \; m = -1, \\
 0 & \text {for} \; m \in \{0, \ldots, \tau-1 \}, \\
 1 & \text {if} \; m = \tau.
\end{cases}
\end{equation}
This means that for all times $k=j-1, \ldots, j+\tau$, the observed state
vectors $\mathbf{v}_k$ are compared with $\mathbf{v}_i$. The structure
given by Eq.~(\ref{wvl_eq}) can be interpreted as follows: At time $k=j-1$, the
trajectory falls into an $\varepsilon$-neighborhood of $\mathbf{v}_i$. Then, for
$k=j, \ldots, j+\tau-1$, it moves further away from $\mathbf{v}_i$ than a
distance $\varepsilon$, until it returns to the $\varepsilon$-neighborhood of
$\mathbf{v}_i$ again at $k=j+\tau$. Hence, given a uniform sampling of the
trajectory in the time domain, the length $\tau$ of the resulting white line in
the corresponding RP is proportional to the time that the trajectory needs to
return $\varepsilon$-close to $\mathbf{v}_i$. Going beyond the concept of
\textit{first-return times}, the ensemble of all recurrences to the
$\varepsilon$-neighborhood of $\mathbf{v}_i$ induces a RT distribution for this
specific point. Combining this information for all available points
$\mathbf{v}_i$ in a given time series (i.e., considering the lengths of all
white lines in the RP), one obtains the RT distribution $p(\tau)$ associated
with the observed (sampled) trajectory segment in phase space. Hence, the length
distribution $p(l)$ of white vertical lines $l$ in the RM not containing any
recurrent pair of observed state vectors provides an empirical estimate of the
distribution of RTs on the considered orbit.

\subsection{Recurrence network analysis}

Recently, the idea of transforming a time series into complex network
representations has emerged in the scientific literature, providing new
alternatives for studying basic properties of time series from a complex network
perspective
\cite{Zhang2006,Xu2008,Lacasa2008,Marwan2009,Donner2010NJP,Donner2011IJBC}.
Among other corresponding approaches, the RM (Eq.~\ref{eq_defrp}) can be
re-interpreted as the adjacency matrix of the so-called $\varepsilon$-recurrence
network (RN).

In order to construct the RN, we re-consider the recurrence matrix $R_{i,j}$,
the main diagonal of which is removed for convenience, as the adjacency matrix
$A_{i,j}$ of an undirected complex network associated with the recorded
trajectory, i.e.,
\begin{equation}
A_{i,j}=R_{i,j}(\varepsilon)-\delta_{i,j},
\end{equation}
where $\delta_{i,j}$ is the Kronecker delta. The nodes of this network are
given by the individual sampled state vectors on the trajectory, whereas the
connectivity is established according to their mutual closeness in phase space.
This definition provides a generic way for analyzing phase
space properties of complex systems in terms of RN topology
\cite{Donner2010NJP,Donner2011EPJB}. However, since this topology is
invariant under permutations of the order of nodes, the statistical properties of RNs do
not specifically capture the system's dynamics, but its \textit{geometric} structure
based on an appropriate sampling. We emphasize that a
single finite-time trajectory does not necessarily represent the typical
long-term behavior of the underlying system. Hence, the resulting network properties
can depend -- among others -- on the length $N$ of the considered time series (i.e.,
the network size), the probability distribution of the data, embedding
\cite{Donner2010PRE}, sampling \cite{Facchini_pre_2007,Donner2011IJBC}, etc.

Although they primarily describe geometric aspects, the topological features of
RNs are closely related to dynamical characteristics of the underlying
system \cite{Donner2010NJP,Donner2011EPJB}. In dissipative chaotic model systems (e.g., R\"ossler
and Lorenz systems), both local and global network properties have already been
studied in great detail \cite{Donner2011EPJB}.

In this paper, we consider the following three
characteristics~\cite{Newman2003,Boccaletti2006} as potential candidates for discriminatory statistics:
\begin{enumerate}

	\item the \textit{average path length} $\mathcal{L}$, which quantifies the
	average geodesic (graph) distance $l_{i,j}$ between all pairs of nodes $(i,
	j)$,
	\begin{equation}
		\mathcal{L} = \left<l_{i,j} \right> = \frac{2}{N(N-1)} \sum_{i<j}{l_{i,j}},
	\end{equation}
  where $l_{i,j}$ is the minimum number of edges separating two nodes $i$ and $j$;

  \item the \textit{global clustering coefficient}
	  $\mathcal{C}$ \cite{Watts1998}, which gives the arithmetic mean of
	  the local clustering coefficients $\mathcal{C}_i$ (i.e., the fraction of
	  nodes connected with a node $i$ that are pairwise connected themselves) taken
	  over all $i$,
	  \begin{equation}
 		\mathcal{C} = \frac{1}{N}\sum_{i=1}^{N}\mathcal{C}_{i}
	  \end{equation}
		\noindent
		with
		\begin{equation}
		\quad
		\mathcal{C}_i=\frac{\sum_{j,k; i\neq j\neq k} A_{i,j}A_{i,k}A_{j,k}}{\sum_{j,k; i\neq j\neq k} A_{i,j}A_{i,k}};
		\end{equation}

	\item \textit{network transitivity} $\mathcal{T}$~\cite{Barrat2000,Newman2001},
	which is closely related to $\mathcal{C}$ (but gives less weight to poorly
	connected nodes~\cite{Donner2011EPJB}) and globally characterizes the
	linkage relationships among triples of nodes in a complex network (i.e., the
	probability of a third edge within a set of three nodes given that the two
	other edges are already known to exist),
	\begin{equation}	
		\mathcal{T} = \frac{3\ N_{\Delta}}{N_{3}}
		=\frac{\sum_{i,j,k; i\neq j\neq k} A_{i,j}A_{i,k}A_{j,k}}{\sum_{i,j,k; i\neq j\neq k} A_{i,j}A_{i,k}},
	\end{equation}
	where $N_{\Delta}$ is the number of triangles in the network and $N_{3}$ is the
	number of connected triples. Note that $\mathcal{T}$ is
	sometimes referred to as the (Barrat-Weigt) global clustering coefficient,
	often also denoted as $\mathcal{C}$, e.g., in ref.~\cite{Zou2010}. In order to avoid
	confusion, in this work we prefer to discuss both measures separately.

\end{enumerate}

We emphasize that further network measures (e.g., local betweenness centrality
$b_v$ and global assortativity coefficient $\mathcal{R}$) have also been shown
to discriminate between different types of dynamics in dissipative
systems~\cite{Marwan2009,Donner2010NJP,Zou2010}, but are not
considered in this work for brevity.


%

\end{document}